\documentclass[useAMS,usenatbib,usegraphicx]{mn2e}
\usepackage{graphics}
\usepackage{amssymb}
\usepackage{amsfonts}
\usepackage{wasysym}
\usepackage{epsfig}

\usepackage{longtable}
\usepackage{times}
\usepackage{mathptmx}
\usepackage[usenames,dvipsnames]{color}

\newcommand{\m}{\mathrm}

\title[Modeling of the stellar auroral radio emission] 
{
3D-modelling of the stellar auroral radio emission 
}
\author[P. Leto et al.]
{P. Leto$^{1}$ \thanks{E-mail: pleto@oact.inaf.it},
C. Trigilio$^{1}$,
C. S. Buemi$^{1}$,
G. Umana$^{1}$,
A.~Ingallinera$^{1}$,
L. Cerrigone$^{2}$
\\
$^{1}$INAF - Osservatorio Astrofisico di Catania, Via S. Sofia 78, 95123 Catania, Italy\\
$^2$ASTRON, the Netherlands Institute for Radioastronomy, PO Box 2, 7990 AA Dwingeloo,The Netherlands
}
\begin{document}

\date{}

\pagerange{\pageref{firstpage}--\pageref{lastpage}} \pubyear{}

\maketitle

\label{firstpage}

\begin{abstract}
The electron cyclotron maser is the coherent emission process 
that gives rise to the radio lighthouse effect observed
in the hot magnetic chemically peculiar star CU\,Virginis.
It has also been proposed to explain the highly circularly polarized
radio pulses 
observed on some ultra cool dwarfs, with spectral type earlier than M7.
Such kind of coherent events resemble
the auroral radio emission from the
magnetized planets of the solar system.
 In this paper, we present a tridimensional 
model able to simulate the timing and profile 
of the pulses
emitted by those stars characterized by a dipolar magnetic field by following the hypothesis 
of the laminar source model, used to explain the beaming of the terrestrial auroral
kilometric radiation.
This model proves to be a powerful tool to understand the auroral radio-emission  phenomenon, 
allowing us to derive some general conclusions about the
effects of the model's free parameters on the features of the coherent pulses,
and to learn more about the detectability of such kind of pulsed radio emission.
\end{abstract}



\section{Introduction}

The Electron Cyclotron Maser (ECM) is a coherent emission mechanism 
excited by the inverse velocity distribution that  an electron population propagating 
along the converging field lines of a density-depleted magnetospheric cavity  
can develop. 
This mechanism \citep{wu_lee79,melrose_dulk82} amplifies the extraordinary magneto-ionic mode,
producing almost 100\% circularly polarized radiation at frequency close to
the local gyro-frequency ($\nu_{\mathrm B} \propto B$).
%
%
%
The maser amplification can occur within the magnetospheric region covering a wide range of magnetic field strength,
providing observable ECM emission over a wide frequency range.
The ECM is the mechanism responsible for the broad-band auroral radio emission from the magnetized planets of the solar system, 
such as the terrestrial auroral kilometric radiation (AKR), the Jupiter decametre (DAM), hectometre (HOM) and kilometre (KOM) emission,
the Saturn kilometric radiation (SKR), the Uranus kilometric radiation (UKR) and the Neptunian kilometric radiation (NKR) 
\citep{zarka98}.
The angular beaming of the Earth's AKR, observed with the spacecraft array Cluster \citep*{mutel_etal08},
confirms that the Earth ECM emission pattern is strongly anisotropic. Such terrestrial coherent emission is 
confined within a narrow beam 
tangentially directed along the walls of
the auroral cavity,
where the amplification mechanism occurs. 
{In the framework of the laminar source model
\citep{louarn_lequeau96a,louarn_lequeau96b} such anisotropy of the ECM radiation pattern is
a direct consequence of the small thickness of the cavity. }
In fact,
the radiation path in the direction perpendicular to the cavity walls is too short
to give significant amplification. 
The AKR will then be refracted upwards by the highest-density regions, 
when it leaves the cavity \citep{mutel_etal08,menietti_etal11}.

The auroral radio emission has also been observed in stars. In particular, such a phenomenon has been
well studied in the case of CU\,Vir, 
a magnetic chemically peculiar (MCP) star. 
The MCPs are early type main sequence stars, characterized by strong magnetic fields 
with a mainly dipolar topology, whose axis is tilted with respect to the
rotational axis (oblique rotator). 
Non-thermal incoherent radio emission from MCPs has been detected in almost 25\% of cases
\citep{drake_etal87,linsky_etal92,leone_etal94}.
In accordance with the oblique rotator model (ORM), the radio emission is also periodically
variable as a consequence of the stellar rotation \citep{leone91,leone_umana93},
suggesting that the radio emission arises from a stable rigidly co-rotating magnetosphere (RRM).
This variability has been successfully reproduced by a
3D model able to compute the gyrosynchrotron emission from a RRM in the framework of the ORM \citep{trigilio_etal04,leto_etal06}.

The radio emission from the MCPs is ascribed to a radiatively-driven stellar wind.
Far from the star, specifically on the magnetic equator  at about the Alfven radius, this wind opens the magnetic field lines, forming current sheets.
The reconnection of
the low-strength magnetic field lines accelerates electrons up to relativistic energy.
The thin transitional magnetospheric layer between the inner and dense magnetospheric regions
(the region where the confined plasma accumulates) and the escaping wind is named `middle magnetosphere'. 
The energetic electrons that recirculate through this layer back 
to the stellar surface radiate radio waves by incoherent gyrosynchrotron emission mechanism.
The middle magnetosphere 
is also the region where the conditions for triggering the ECM mechanism could be realised.
In fact, the relativistic electron population,
accelerated 
far from the star,
moves along the converging dipolar field lines
toward the high-strength magnetic field regions close to the stellar surface.
Moreover, due to magnetic mirroring, only energetic electrons with very low pitch-angle (electron path
almost parallel to the magnetic field lines) can impact the stellar surface and will be lost.
The reflected non-thermal electron population
could then develop the inversion of the velocity distribution
needed to trigger the ECM mechanism.
Other kinds of unstable energy distribution can be developed by the precipitating 
non-thermal electron beams, such as the horseshoe distribution, able to efficiently produce ECM emission \citep{ergun_etal00}.

In spite of the favourable physical conditions that characterize the magnetospheres of the radio MCP stars, 
 only CU\,Vir and HD\,133880 are so far characterized by broad-band, 
highly polarized and time-stable pulses \citep{trigilio_etal00, trigilio_etal08, trigilio_etal11, ravi_etal10, lo_etal12,chandra_etal15}, 
ascribed to auroral radio emission
detected when the magnetic dipole axis is almost perpendicular to the line of sight.
The frequency of the ECM emission observed on CU\,Vir ranges from about 600 MHz \citep{stevens_george10} to 5 GHz 
(tentative detection reported by \citealp{leto_etal06}), whereas for HD\,133880
a detection of ECM emission at 600 MHz and 1.4 GHz  has been reported \citep{chandra_etal15}. 

CU\,Vir is the first star where the stellar auroral radio emission was detected \citep{trigilio_etal00}.
On this star, the coherent emission process was
mainly observed and extensively studied at the frequencies of 1.4 and 2.5 GHz
\citep{trigilio_etal00,trigilio_etal08,trigilio_etal11,ravi_etal10,lo_etal12}. 
It shows two pulses per spin period, almost 100\% right hand circularly polarized, 
with a phase separation of $\approx 0.4$, and 
 characterized by a frequency drift of the pulse arrival time.
Two pulses per period have been detected at 1.4 GHz in every observing epochs,
unlike the auroral radio light curve at 2.5 GHz, which does not always show two pulses  per period.
Moreover, 
left-handed circularly polarized pulses were never detected from CU\,Vir.
This lack of left-hand polarization has been explained as due to the non-perfectly dipolar topology of its magnetic field \citep{kochukhov_etal14}. 
The existence of multipolar magnetic-field components makes
the behaviour of the two opposite hemispheres asymmetric. 
In addition, the existence of a multipolar field topology has been indicated as the cause for the absence
of  auroral radio emission in $\sigma$\,Ori\,E \citep{leto_etal12}, another well-studied radio MCP star 
characterized by a non-simple dipole field \citep{oksala_etal12}.

Similarly to the Earth AKR, the coherent pulses observed from CU\,Vir
were explained as ECM emission originating in an auroral cavity,
tangentially beamed to the walls and then upward refracted \citep{trigilio_etal11}.
This anisotropic beaming
is able to successfully reproduce the timing and the pulse width of the auroral radio emission observed
in CU\,Vir \citep{lo_etal12}.

At the bottom of the main sequence,
coherent pulses, still explained as auroral radio emission due to the ECM, were also
observed in some ultra cool dwarfs (UCDs) with spectral type earlier than M7
\citep{berger02, 
burgasser_putman05, 
antonova_etal08, 
hallinan_etal08, 
route_wolszczan12, 
route_wolszczan13, 
williams_etal15}, 
showing in some cases features similar to those observed in CU\,Vir. For example,
the M8 dwarf DENIS-P\,J1048.0\,--\,3956 shows time-shifted pulses at 5 and 8.4 GHz,
both fully right-hand circularly polarized \citep{burgasser_putman05} and the T6.5 brown dwarf 2MASSI\,J1047539\,+\,212423 is characterized by left-hand circularly polarized periodic pulses
detected at 6 GHz \citep{williams_etal15}.

In the framework of the 
laminar source model,
in this paper we develop a 3D-model 
of the auroral radio emission from stars with a dipole-like magnetic field.
This model is used to study how the auroral radio-emission features depend on the stellar geometry and on
the  parameters that define the ECM beam pattern.
This study was able to give a deeper insight about the relationship between the auroral radio emission detectability and the model parameters.


\begin{figure}
\resizebox{\hsize}{!}{\includegraphics{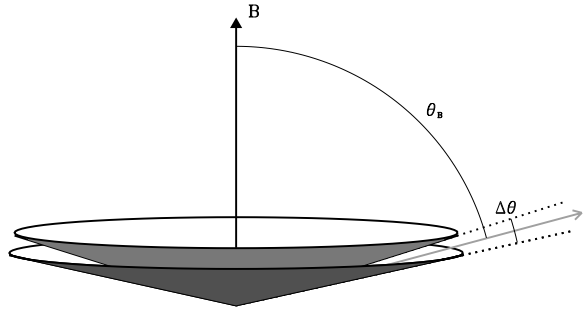}}
\caption{{Beaming pattern of the loss-cone driven ECM  elementary emission process.
The amplified radiation, displayed by the light-grey vector, is emitted within an hollow-cone
centered on the local magnetic field vector ($B$) of
half-aperture $\theta_{\mathrm B}$, and thickness $\Delta {\theta}$.}
}
\label{geo}
\end{figure}

\section{The auroral radio emission model}

{For the ECM emission driven by the loss-cone instability, the radiation amplified by
this kind of coherent emission process is beamed in a very thin hollow cone
centered on the local magnetic field line \citep{wu_lee79,melrose_dulk82}.
The radiation cone have
half-aperture         
$\theta_\mathrm{B} \approx \arccos (v/c )$, where $v$ is the speed of the unstable electrons population
\citep*{hess_etal08}.
In the case of  2 KeV energy electrons the hollow cone half-aperture is $\approx 85^{\circ}$.
To well explain the geometry of the elementary emitting process, 
a thin conical sheet of thickness $\Delta\theta$ has been shown in Fig.~\ref{geo}.
The frequency $\nu$ of the amplified radiation has to satisfy the cyclotron resonance condition
for transverse emission 
$\nu = s \nu_{\mathrm B} / \gamma$, where $s$ is  the harmonic  number
of the local gyro-frequency ($\nu_{\mathrm B}=2.8 \times 10^{-3} B/{\m G}$ GHz), 
for mildly relativistic electrons the Lorentz factor $\gamma \approx 1$.
The fundamental harmonic ($s=1$) has
the fastest growth rate, but it is likely suppressed by the gyromagnetic absorption as it crosses the more external layer 
in which the second harmonic of the local gyro-frequency is equal to the amplified frequency.
The layers at $s>2$ have a lower optical depth for gyromagnetic absorption,
it is thus reasonable to assume that, when detected, the ECM emission occurs at $\nu=2 \nu_{\mathrm B}$ \citep{melrose_dulk82}.
}


In the case of ECM amplification occurring in thin magnetospheric cavities (laminar source model),
the resulting overall radiation diagram can be strongly anisotropic  \citep{louarn_lequeau96a,louarn_lequeau96b} and
the axial symmetry of the single electron emission beam will be lost. 
In fact, the elementary ECM sources with similarly-oriented emission beam pattern
and located along the line of sight contribute all together  to the maser amplification. 
If the cavity has a laminar structure, the rays tangent to the walls
maximize the path within the region where the ECM originates.
Theoretical simulations also confirm that the auroral radio emission is mainly amplified tangentially to 
the cavity boundary, rather than along the perpendicular direction \citep{speir_etal14}.
The resulting ECM emission is beamed in a strongly directive radiation diagram 
and will then be detectable only when our line of sight crosses the radiation beam pattern,
like the pulses of a radio light-house.

%
%
%
%
%
%

\begin{figure}
\resizebox{\hsize}{!}{\includegraphics{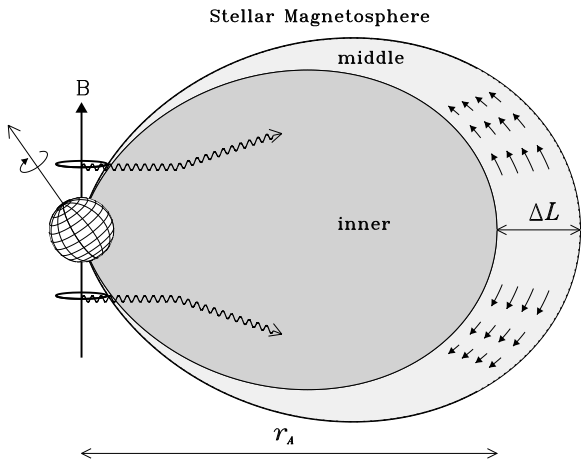}}
\caption{Section of a dipole-like magnetosphere, where the auroral radio emission can take place.
The magnetospheric size is defined by the Alfven radius $r_{\mathrm A}$, whereas the thickness ($\Delta L$) of
the auroral cavity is defined by 
the lengths of the current sheet where the free electrons are accelerated up to relativistic energy.
The non-thermal electrons (represented by the small solid arrows) 
move toward the stellar surface within a transitional layer named middle magnetosphere. 
Above the magnetic poles, close to the stellar surface, the two auroral rings where the ECM arises are located.
The amplified radiation (the two waves displayed in the northern and southern hemispheres)
is emitted tangentially to the ring and then upward refracted by the
dense thermal plasma trapped inside the inner magnetosphere.
}
\label{schema}
\end{figure}

We will analyze the auroral radio emission from a dipole-like magnetospheric cavity in the framework of the
laminar source model described above.
The magnetosphere is assumed as a simple dipole tilted with respect to the rotation axis. Within a fixed value of the equatorial radius
(Alfven radius: $r_{\mathrm A}$), the thermal plasma accumulates (inner magnetosphere), while beyond the Alfven radius
the plasma breaks the magnetic field lines giving rise to a density-depleted magnetospheric cavity (middle magnetosphere),
where the non-thermal electrons can freely propagate, accelerated by magnetic reconnection
(see Fig.~\ref{schema}). This population of
non-thermal electrons can develop an unstable energy distribution,
able to pump the electron cyclotron maser. 
The overall  behavior of the ECM emission process that can take place in 
this kind of magnetospheric cavity
(schematically pictured in Fig.~\ref{schema}) is described as follows:
the ECM radiation arising from a generic point of the auroral ring propagates in the plane of the ring;
following the laminar source model, the coherent emission 
is beamed along the plane parallel to the magnetic axis and tangent to the ring; 
then the auroral radio emission is upward refracted by the dense thermal plasma
trapped within the inner magnetosphere.
%
%
%
%
%
To study the  features of the ECM pulses arising from such kind of auroral cavity, 
we use the same approach 
followed 
by \citet{trigilio_etal04} and \citet{leto_etal06}.
In these works, we simulated the  incoherent gyrosynchrotron brightness distribution and the total flux density 
by sampling the space surrounding the star. 
Given physical parameters such us the local magnetic field strength
and orientation, the thermal and non-thermal electron number density etc., we calculated at every grid point
the local emission and absorption coefficients for the gyrosynchrotron mechanism, which are needed to integrate the radiative transfer equation 
along the line of sight.
In the present work, we do not integrate the radiative transfer equation for the ECM.
We only search for those spatial points that satisfy  the physical conditions 
required for the electron cyclotron maser generation at a given frequency and that have
the radiation beam pattern oriented along the line of sight.
The modelling approach developed in this paper is only limited to the study of the timing and profile 
of  the ECM pulses, which depends on those parameters that
define the geometric conditions able to make such phenomenon detectable. 

As a first step, we set the stellar geometry ($i$ is the inclination of the rotation axis, $\beta$ the tilt of the dipole magnetic axis,
and $\Phi$ the rotation phase)
and the polar field strength. The rotation axis, which is misaligned with respect to the magnetic polar axis, is displayed in Fig.~\ref{schema}.
In the stellar reference frame, the space surrounding the star is sampled in a 3D  cartesian grid and  in each grid point
the dipolar magnetic field vector components are calculated.
The vectorial field is then rotated in the observer's reference frame (see appendix A of \citealp{trigilio_etal04}).


As a second step, we localise the magnetospheric cavity where the unstable electron
population propagates. 
The inner and outer boundary dipolar field lines are defined by the equation: $r=L \cos^2 \lambda$, where
$r$ is the distance to the centre of the star and $\lambda$ the magnetic latitude.
This space region intercepts the magnetic equatorial plane
at the distances  $L$ and $L+\Delta L$ from the centre of the star, respectively for the inner and outer boundary.
These magnetic field lines are shown in Fig.~\ref{schema}.
Within the magnetic shell thus defined,
we can find the set of grid points that have the same value of the local magnetic field strength.
Given the frequency $\nu$ of the observable coherent emission
we are so able to localize those grid points that have $2 \nu_{\mathrm B}=\nu$, within a fixed bandwidth $\Delta \nu$.
These points are distributed in polar rings above the magnetic poles.
The northern and  southern auroral rings that emit the ECM at an arbitrary frequency
are displayed in Fig.~\ref{schema}. 
Given the boundary magnetic field lines,
the radius and height above the stellar surface of these rings 
follow the radial dependence of the magnetic field strength $B$, which defines the gyro-frequency.

\begin{figure}
\resizebox{\hsize}{!}{\includegraphics{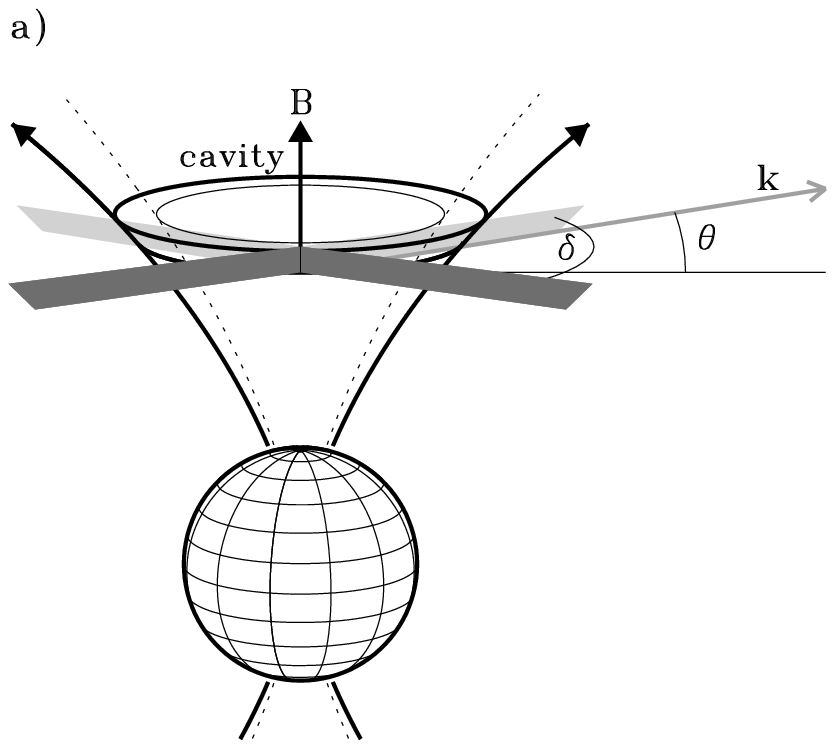}}
\resizebox{\hsize}{!}{\includegraphics{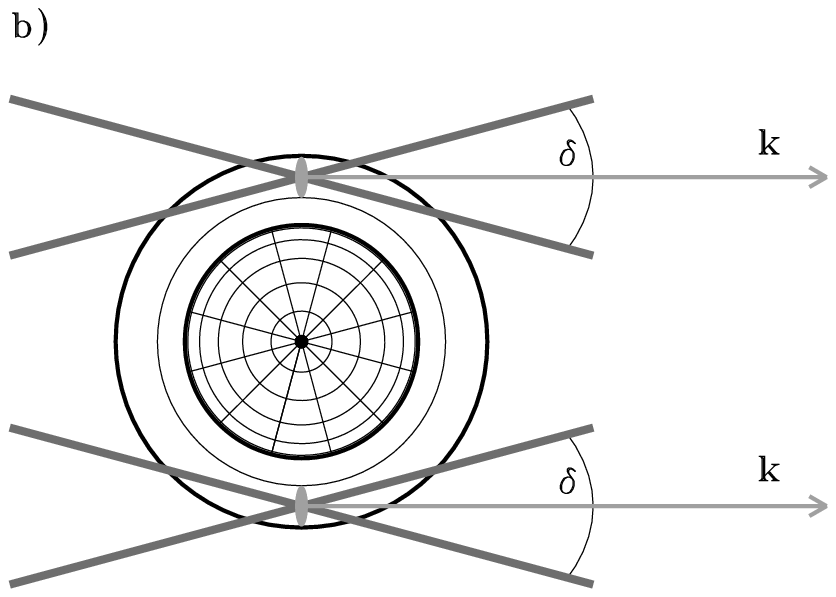}}
\caption{
Schematic view of  the model of the auroral radio emission developed in the framework of the 
laminar source model.
Panel a) side view; panel b) top view. 
The density-depleted magnetic cavity and the boundary field lines are shown.
A representative auroral ring is also displayed above the north pole, where the ECM emission process originates.
The grey planes passing trough the boundary of the cavity describe the ECM beam pattern, whose size is given by the angle $\delta$.
The ECM ray vector ($k$) is misaligned of an angle $\theta$ with respect to the direction perpendicular to the local magnetic field vector ($B$).
}
\label{tp}
\end{figure}

The third step consists in the definition of
the emission beam pattern of each grid point belonging to the auroral rings.
For clarity, the side and top view of the ECM emission diagram arising from the northern ring
are displayed in the two panels of Fig.~\ref{tp}.
Each point of the auroral ring, which is the source of ECM emission at frequency $\nu$, 
has an emission diagram defined by the following angles: 
{the deflection angle $\theta$; the opening angle $\Delta \theta$; the beaming angle $\delta$.
The angle $\theta$ is defined as the angle between the ray path vector $\mathbf{k}$ and 
the plane containing the ring (that is perpendicular to the magnetic dipole axis).
$\Delta \theta$, set equal to 5$^{\circ}$,
is the hollow-cone thickness (see Fig.~\ref{geo}) that in the case of the loss-cone driven instability is a
function of the emitting electrons speed as  $\Delta \theta \approx v/c$ \citep{melrose_dulk82}, and
in accord with \citet{trigilio_etal00} such speed has been fixed equal to $0.09c$
(electrons energy $\approx 2$ KeV), corresponding to an emitting conical sheet just 5$^{\circ}$ thick.
The angle $\delta$ accounts for the width of the radiation diagram centered on the plane 
tangent to the auroral circle and parallel to the magnetic axis.
This angle have the aim to parameterize the number of elementary ECM sources, 
located in the auroral rings,
with the emission beams closely aligned with the line of sight}
(the beam width is highlighted by the grey planes  in Fig.~\ref{tp}). 
When the line of sight passing through a given grid point located on the auroral circle
is tilted with respect to the local magnetic field vector of an angle $90^{\circ}-\theta$
(for ECM sources in the northern magnetic hemisphere) or $90^{\circ}+\theta$
(for sources in the southern hemisphere) {within $\Delta\theta/2$},
and forms an angle lower then $\mid \delta /2 \mid$ with
the plane tangent to the auroral ring, 
the corresponding grid point will be assumed as an observable source of ECM emission.

To separate the contributions of the ECM emission arising from the two opposite magnetic hemispheres,
we assign the right-hand circular polarization (RCP) to the emission from sources
 in the northern hemisphere and the
left-hand circular polarization (LCP) to the emission from the southern ones.
If we repeat the operations explained above as a function of the stellar rotation phase and
store the number of grid points that are ECM sources, along with their corresponding polarization sense, then we are 
 able to simulate the features of the ECM light curves for the I and V Stokes parameters
(respectively defined as ${\mathrm {RCP}}+{\mathrm {LCP}}$ and ${\mathrm {RCP}}-{\mathrm {LCP}}$).


As a further output,
once we set the stellar parameters that describe the geometry and strength of the dipolar field,
the 3D-model provides us with the effective magnetic field curve.
In fact, the simulations of the ECM emission require the calculation of the vectorial field generated by the magnetic dipole.
For a given rotational phase, we are able to calculate
the longitudinal component of each magnetic field vector located on the stellar surface.
{The sum of such longitudinal components, weighted by the limb darkening law,
gives us the theoretical effective magnetic field strength ($B_{\mathrm{e}}$)
as a function of the stellar rotation. In detail, such calculation has been performed as follow:
\begin{displaymath}
B_{\mathrm{e}}= \frac{\sum_{{x}}\sum_{{y}}\sum_{{z}} B_{\mathrm{x}}(x,y,z) I(x,y,z)}{\sum_{{x}}\sum_{{y}}\sum_{{z}} I(x,y,z)}
\end{displaymath}
where $B_{\mathrm{x}}$ is the longitudinal component of the magnetic field vector located at the coordinate $(x,y,z)$
in the side of the stellar surface facing to the observer,
and $I$ is the corresponding limb darkening parameter.
The observing reference frame ($Oxyz$) is centered on the stellar center, 
the $x$-axis is aligned with the line of sight and oriented toward the observer,
the $y$ and $z$-axes individuate the plane of the sky.
Each point of the grid is associated to
the vector $\vec{r} \equiv (r_{\mathrm{x}},r_{\mathrm{y}},r_{\mathrm{z}})$, 
and the points located on the stellar surface are those that verify, within the sampling step, the condition
$r_{\mathrm{x}}^2+r_{\mathrm{y}}^2+r_{\mathrm{z}}^2=1$ (assuming a unitary stellar radius). 
For the calculation has been adopted the simple linear limb darkening law:
\begin{displaymath}
I(x,y,z)=1-k + k{r_{\mathrm{x}}(x,y,z)} ~~~~~~~~(0\leq {r_{\mathrm{x}}} \leq 1)
\end{displaymath}
with $k=0.45$ \citep{stift73}.}
Therefore, with the capability to simultaneously simulate the ECM light curve and the magnetic curve,
we can correlate the ECM occurrence with the  $B_{\mathrm{e}}$ curve.


\begin{figure}
\resizebox{\hsize}{!}{\includegraphics{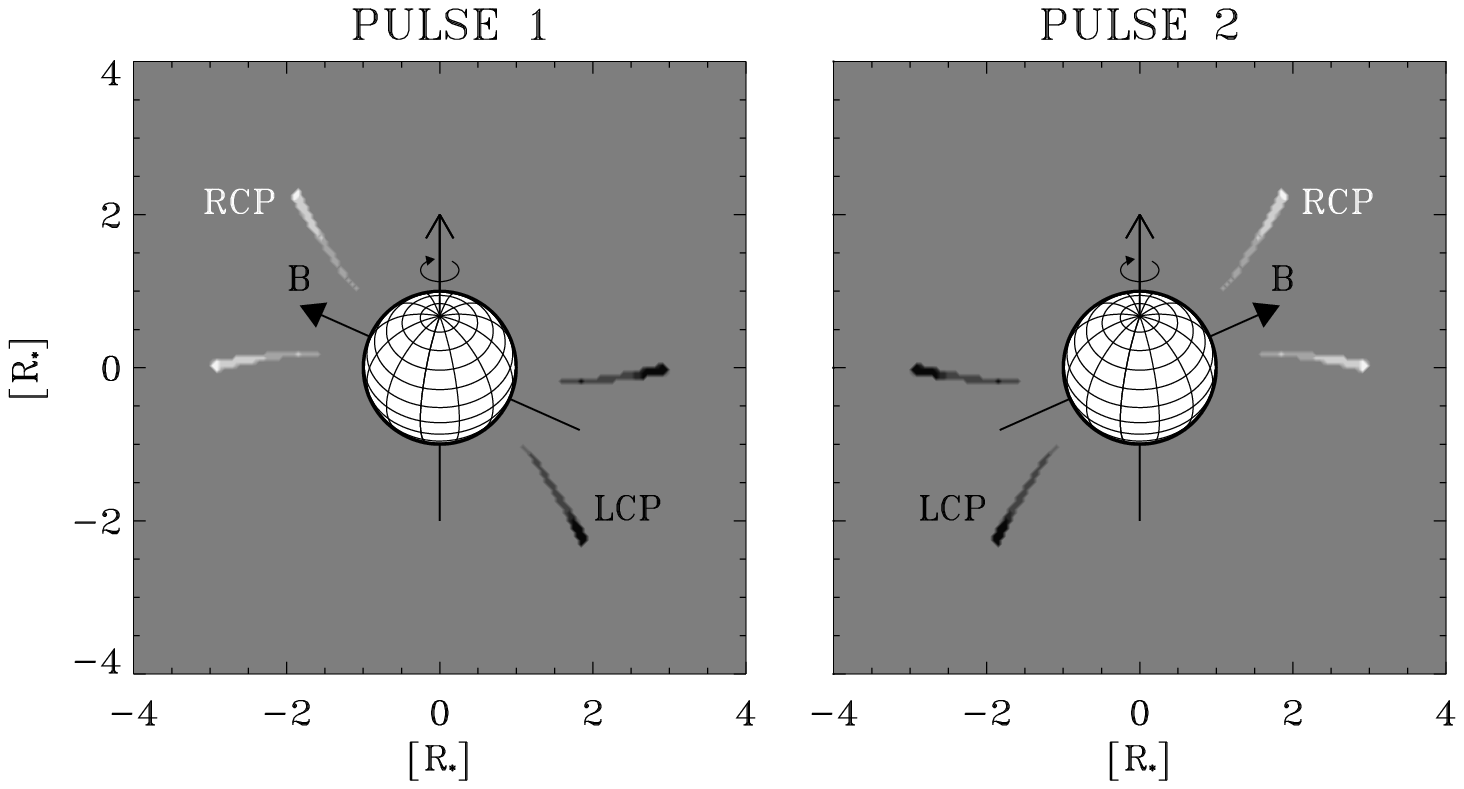}}
\resizebox{\hsize}{!}{\includegraphics{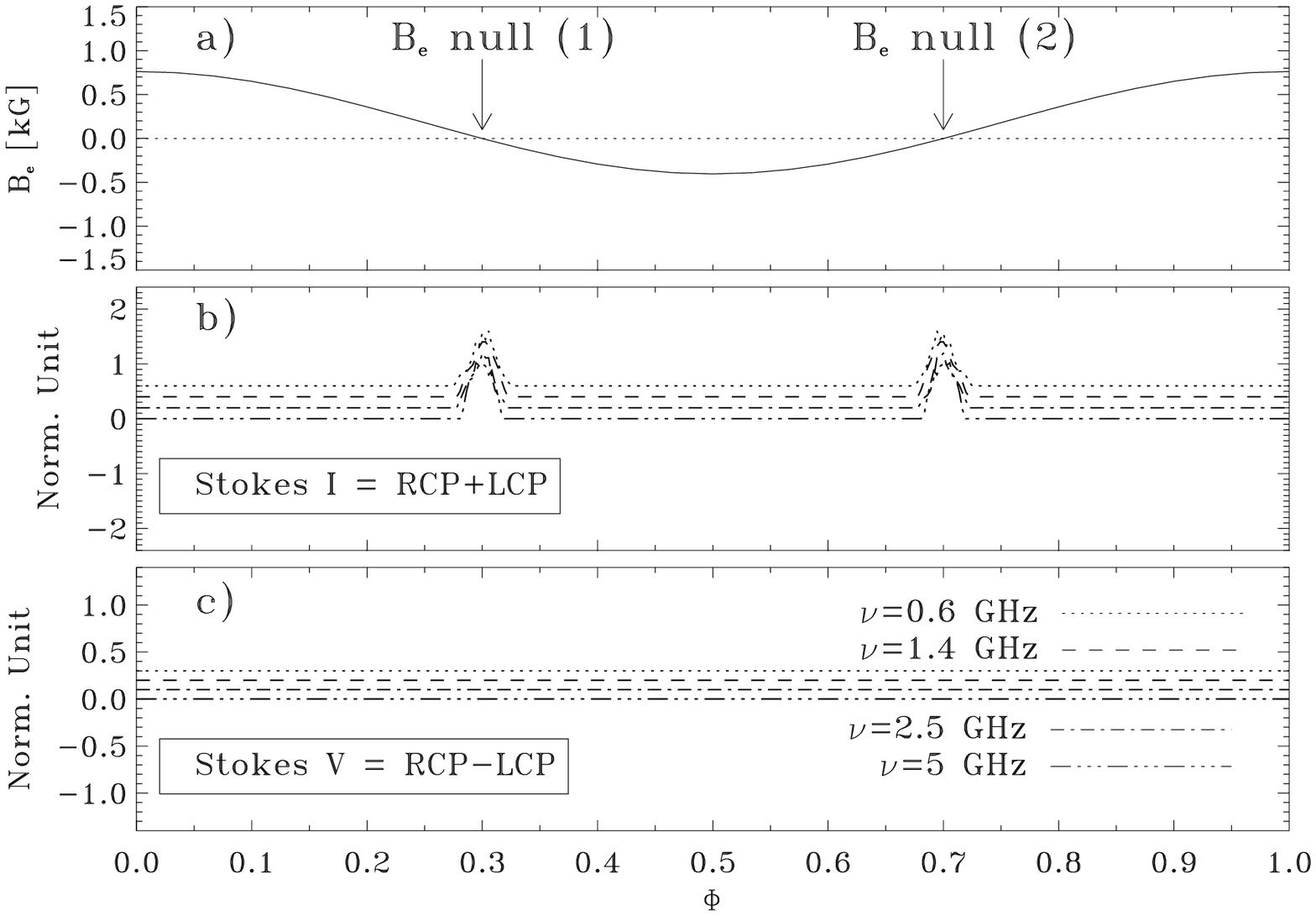}}
\caption{
Top panels: wide band
synthetic brightness spatial distribution of the ECM emission arising from the two
opposite hemispheres of a dipole-like oblique rotator. 
The maps
have been created for
the two stellar orientations where the {longitudinal} magnetic field is null. 
The simulated frequencies range between 5 GHz, close stellar regions, 
and 600 MHz, farther regions. 
The emission that originates from the northern hemisphere is shown by the white area,
while the dark area is associated to the ECM emission from the southern hemisphere.
Bottom panel a): simulated magnetic curve. Bottom panel b) and c): {ECM light curves simulated at $\nu=0.6$, 1.4, 2.5 and 5 GHz, respectively for Stokes I and V. An arbitrary offset has been added to each individual light curve to distinguish them}.
}
\label{mappe2}
\end{figure}

\section{Model Predictions}

\subsection{Dependence on the auroral ring geometry}
\label{sec3_1}

First of all we analyze if the spatial location and the size of the auroral regions that are sources to the
ECM emission affect the simulated light curves. 
To do this, we set the magnetic field and vary the frequency;
this is equivalent to simulate ECM emission from auroral rings of different sizes and locations above the stellar surface.

We simulated  a set of frequencies ranging from 600 MHz to 5 GHz.
For this specific analysis the stellar geometry and the ECM radiation diagram are fixed.
We assumed a dipole with a polar field strength of 3000 gauss, misaligned with respect to the rotation axis
of an angle  $\beta=74^{\circ}$, the rotation axis inclination ($i$) has been fixed to $43^{\circ}$.
The parameters above are equal to the stellar
parameters adopted for CU\,Vir (the prototype of the stars showing auroral radio emission) by \citet{trigilio_etal00}.
The thickness of the magnetic shell that delimits the auroral cavity has been defined
setting the $L$-shell parameter of the two boundary dipolar magnetic field lines, respectively equal to 15 and 18 R$_{\ast}$ ($\Delta L=20\%L$),
in accordance with the size of the magnetospheric region, where the incoherent radio emission of CU\,Vir originates \citep{leto_etal06}.
The two boundary magnetic field lines have the maximum separation at the magnetic equator, 
whereas at the stellar surface these two field lines are very close.
For the ECM emission beam pattern, we assume 
the simple condition of amplified radiation propagating perpendicularly
to the local magnetic field vector (no refraction, $\theta=0^{\circ}$) with a narrow radiation diagram
(opening angle fixed  to $\delta=10^{\circ}$). 

The spatial distribution of the ECM emission, 
calculated when the magnetic dipole axis lies exactly in the sky plane (null value of the effective magnetic field),
has been displayed in the top panels of Fig.~\ref{mappe2}.
The simulated maps were obtained by collecting the ECM contributions
from all grid points (simulation step equal to 0.05 R$_{\ast}$) that have the second harmonic of the local
gyrofrequency in the range from 600 MHz to 5 GHz. 

In accordance with the radial dependence of the magnetic field strength, decreasing outward as $r^{-3}$ for a dipole,
the highest frequencies originate in small auroral rings located near the star,
the low frequencies are instead generated in large auroral rings at high distance above the stellar surface.
The simulated frequencies originate from auroral rings located between $\approx0.5$ and 2 R$_{\ast}$ from the stellar surface.
The ECM emission at 5 GHz is generated within the thin auroral cavity located close to the star,
unlike the ECM radiation at 600 MHz, which arises from the auroral regions far from the star, and is
consequently generated within a larger and thicker auroral cavity.


\begin{figure*}
\resizebox{\hsize}{!}{\includegraphics{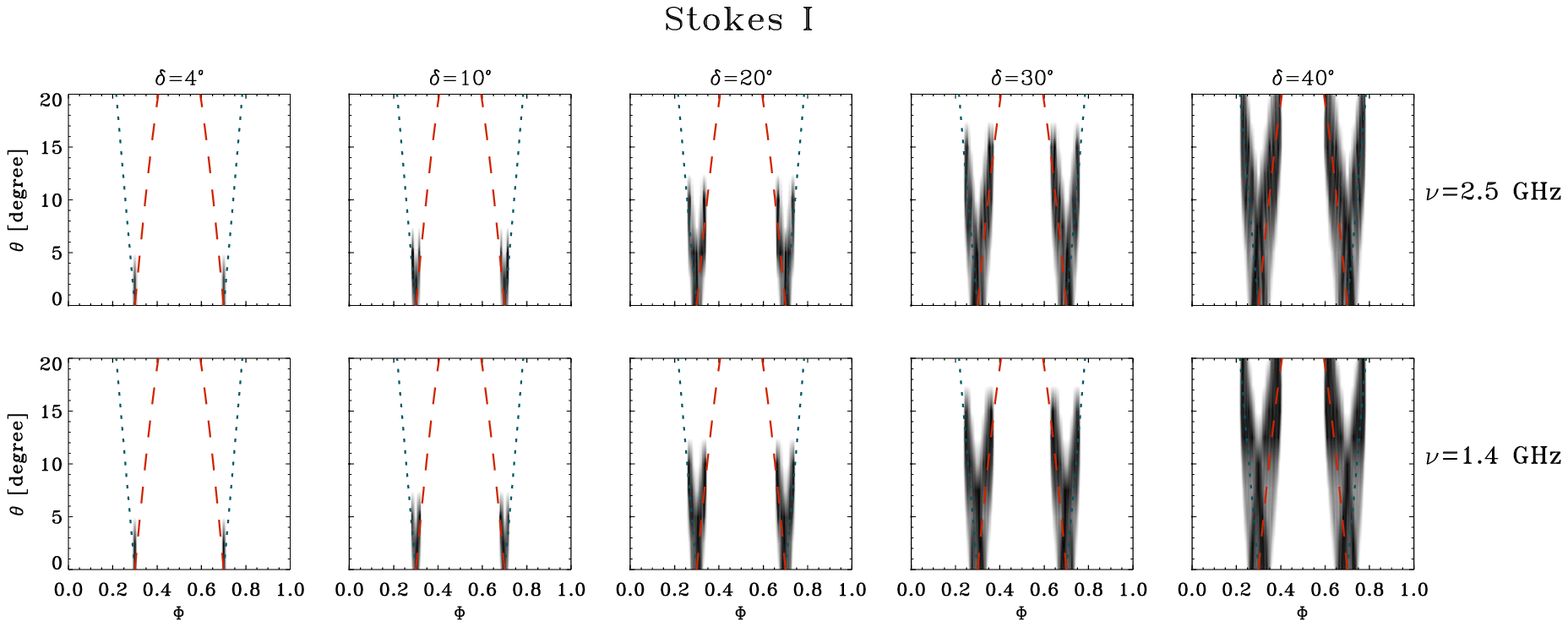}}
\resizebox{\hsize}{!}{\includegraphics{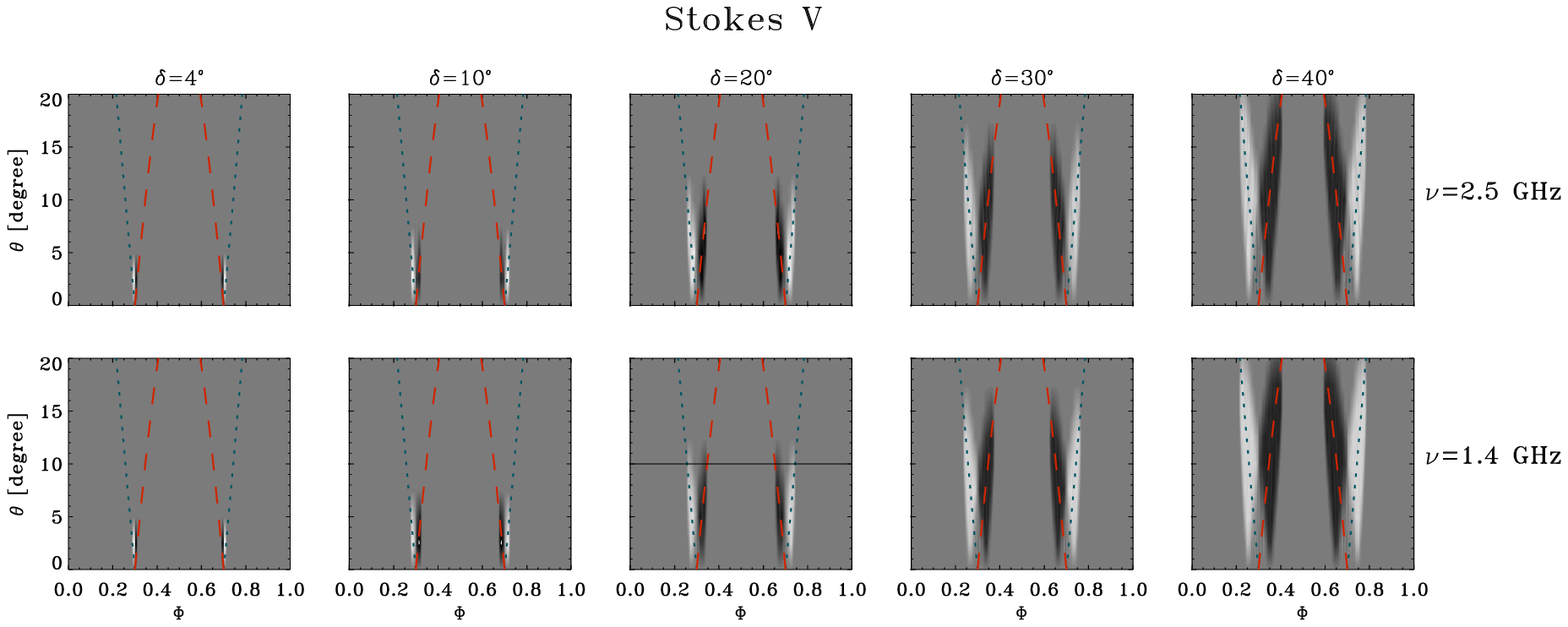}}
\caption{
Dynamic ECM light curves results of the 3D model simulations obtained varying the parameters that control the radiation beam pattern.
The simulated frequencies are: 1.4 and 2.5 GHz. 
The model simulations have been performed as a function of the ray path deflection $\theta$, y-axis of each panel.
The simulations have been performed for 5 values of the opening angle $\delta$.
The top panels display the Stokes I parameter, while the bottom panels display Stokes V; black corresponds to the left handed circularly polarized component,
white indicates right hand polarized.
The pulse phase location predicted by the ORM have also been shown; the blue dotted line is 
related to the ECM pulses arising from the northern hemisphere, the red dashed line is referred to the southern.
The black continuous line in the panel with the simulations performed at $\nu=1.4$ GHz and
$\delta=20^{\circ}$ is a representative light curve obtained assuming $\theta=10^{\circ}$.
}
\label{sim}
\end{figure*}

\begin{figure*}
\includegraphics[width=150mm]{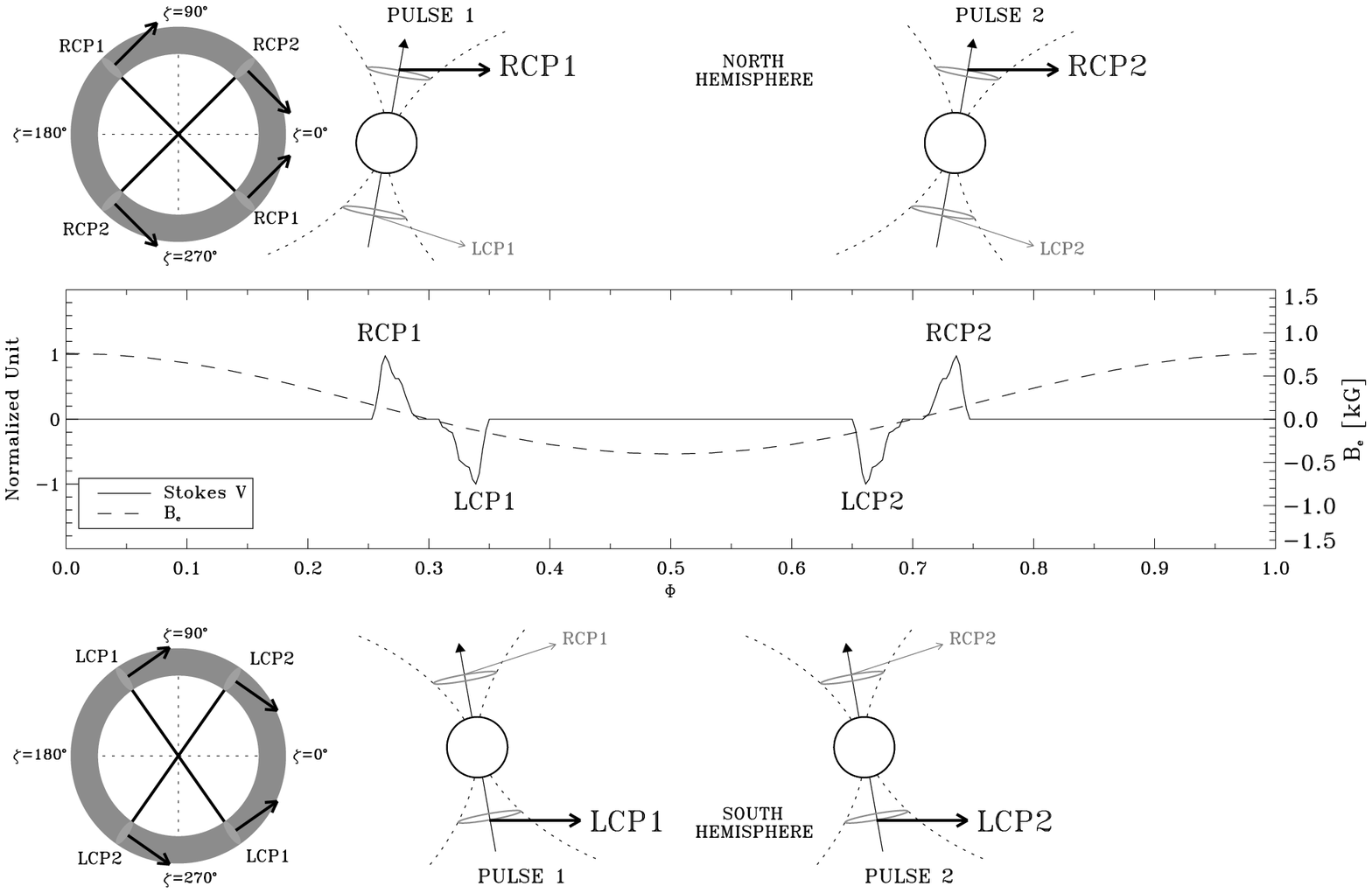}\\
\caption{Synthetic auroral radio light curve (obtained assuming: $\nu=1.4$ GHz, $\delta=20^{\circ}$ and $\theta=10^{\circ}$),
solid line, with superimposed the simulated effective magnetic field curve ($B_{\mathrm p}=3000$ Gauss, $i=47^{\circ}$, $\beta=74^{\circ}$),
dashed line.
The pulses have been labelled as in Fig.~\ref{mappe2} according to the arrival order and the circular polarization signs. 
The cartoons show the magnetosphere orientations corresponding to the beam of the auroral radio emission
aligned with the  line of sight.
The two unit circles indicate the magnetic longitudes ($\zeta$) of the magnetic field lines responsible
of the auroral radio emission observable from Earth.
}
\label{ecm_b}
\end{figure*}

{The model predicts similar ECM light curves at  different simulated frequencies} (small differences 
are due to the sampling effect) characterized by two peaks per stellar rotation, visible when $B_{\mathrm e}=0$.
The theoretical magnetic curve and the corresponding
ECM light curves for the Stokes I and  V are displayed in the three bottom panels of Fig.~\ref{mappe2},
simulated setting {$\nu=0.6$, 1.4, 2.5 and 5 GHz}.
As a consequence of the chosen stellar geometry (assumed equal to the prototype CU\,Vir),
the minimum phase difference between the two simulated ECM peaks is about 0.4,
that is in accordance with the phase difference between the two coherent events observed on CU\,Vir.

By inspecting  the simulated light curves for the two Stokes parameters, we can notice that there is not
any Stokes V (${\mathrm {RCP}}-{\mathrm {LCP}}$) contribution,
whereas the Stokes I (${\mathrm {RCP}}+{\mathrm {LCP}}$) is well observable. 
This is because the ECM contributions from the two opposite
magnetic hemispheres are exactly the same, 
as it is evident by the perfect symmetry of the ECM brightness spatial distribution (top panel of Fig.~\ref{mappe2}).
Since the ECM radiation arising from the two hemispheres has opposite polarization senses in the north and in the south,
the Stokes V parameter will be consequently null.

\subsection{Dependence on the beam pattern}

To assess the effect of the free parameters that control the beam pattern of the
ECM radiation on the auroral radio emission features, 
we simulated the light curves varying the opening angle $\delta$, 
and the upward deflection angle $\theta$, keeping the stellar geometry previously adopted.
In this case, we performed simulations of the ECM emission at 2.5 and 1.4 GHz.
{To mitigates the possible bandwidth effect that could affect the comparison between the simulations and most of
the available observations, these simulations have been performed within a 
narrow frequency range, 100 MHz wide, which is close to the old VLA and ATCA bandwidth set-up.}
For the assumed polar field strength ($B_{\mathrm e}=3000$ Gauss), the two auroral rings 
are located respectively at $\approx 0.9$ and 1.3 R$_{\ast}$ above the magnetic poles.
The results of the simulations for the Stokes parameter I and V are displayed in Fig.~\ref{sim},
where the right circularly polarized emission (positive Stokes V)
is displayed  in white and the left circularly polarization (negative Stokes V) in black. 
In each panel of Fig.~\ref{sim}, for a given angle $\delta$,
the dynamical simulated light curves are displayed as a function of the angle $\theta$,
shown on the $y$-axis.

First of all, we notice that the light-curve features of the auroral radio emission at the two simulated radio frequencies - 1.4 and 2.5 GHz - 
have a closely similar dependence on the two parameters $\delta$ and $\theta$.
This is equivalent to say that the spatial location and the size of the auroral rings have little or no effect
on the main features of the auroral radio emission, once the auroral cavity is defined.
The timing and the pulse profile are instead significantly affected by the choice of $\delta$ and $\theta$.
In particular, the dynamic light curves of the Stokes I parameter show that
the pulse width is directly related to the angle $\delta$, with the narrow peaks associated to the smaller $\delta$ values.
We also note that
the pulses disappear above a deflection angle limit.
This trend is observable for each value of the parameter $\delta$,
with the limit value of the angle $\theta$ 
growing as $\delta$ increases.
It is also evident that, before disappearing, each pulse splits in two, with phase separation growing with  $\theta$.
By analyzing the Stokes V, it is clear that each component of
the double peak pulses has opposite polarization sense.
In general, Stokes V has 
a hybrid polarization, 
as a consequence of the different contributions to the ECM emission of
the two stellar hemispheres with opposite magnetic polarity. 
{As a consequence of the physical processes that are responsible
of the generation and propagation of the ECM emission,
the deflection angle $\theta$ could be
a function of the frequency.
In fact, the basic processes that drive the ECM,
like the loss-cone instability, originate amplified radiation propagating 
in direction which is frequency-dependent, or/and 
the refraction effects,
that can be suffered by the amplified radiation traveling through the cold
thermal plasma layers that are located in the stellar magnetosphere, which are also function of the frequency.
Comparing multi frequency observations and simulations we will be able
to measure the deflection angle of each individual amplified frequency,
and thus we can try to disentangle among the possible frequency-dependent
mechanism that can affect the propagation direction of this kind of amplified emission.}

The pulse timing is the result of the oblique rotator model assumption. In fact, the upward deflection angle $\theta$ 
is related to the angle formed by the line of sight with the magnetic dipole axis ($\theta_{\mathrm M}$) as follows:

~\\
\noindent
$\theta _{\mathrm M}=\left\{
        \begin{array}{lr}
        \pi /2 -\theta & \mbox{Northern Hemisphere} \\
        \pi /2 +\theta & \mbox{Southern Hemisphere}
        \end{array}
\right.
$
~\\

\noindent
Once  the ORM geometry ($i$ and $\beta$) is defined, the phase location of the ECM pulse is expressed by the following equation
\citep{trigilio_etal00}:
\begin{equation}
\Phi_{\mathrm {ECM}}  =  \arccos \left( \frac{\cos \theta _{\mathrm M} - \cos \beta \cos i}{\sin \beta \sin i}  \right) / 2\pi
\label{eq1}
\end{equation}

\noindent
The tracks of the ECM pulses calculated by the use of  Eq.~\ref{eq1} have been put on top of 
the simulated light curves  in Fig.~\ref{sim}.
The comparison between the model simulation and the ORM prediction 
indicates that the phase location of each ECM peak can be predicted by  Eq.~\ref{eq1},
whereas the ECM pulse profile is the result of a complicated combination of the angles
that control the beam of the auroral radio emission.

To better clarify the above model prediction, the 1.4 GHz stokes V synthetic light curve
obtained with $\delta=20^{\circ}$ and $\theta=10^{\circ}$ has been analysed  (it is identified in Fig.~\ref{sim}
by the black line). 
This selected light curve is shown in Fig.~\ref{ecm_b}
along with the stellar magnetic curve.
The stellar orientations associated 
with the various pulse components and
identified on this simulated light curve have also been displayed.
For clarity, each pulse component has been assigned a label related to the order of appearance 
and to the polarization sense. The ECM ray vectors 
have been tagged according to the labels of the corresponding simulated coherent pulse components.

The ECM emission beam pattern 
responsible for the first right-hand circularly polarized peak (RCP1) arising from the northern auroral ring
is aligned with the line of sight at phases slightly preceding the first null of the magnetic curve, depending on $\theta$ (Fig.~\ref{ecm_b}).
As the star rotates, the northern pole will be seen moving away from the observer
and, after the magnetic curve passes the first null,
the ECM emission arising from the southern auroral ring will be visible
(first left-hand circularly polarized pulse: LCP1).
As the star rotates, 
after that the magnetic curve has reached its negative
magnetic outermost point, the magnetic north pole will be seen approaching the observer and
a second ECM beam generated from the southern auroral ring
will be aligned with the line of sight. Consequently, the second LCP pulse will be detectable (LCP2).
Likewise, the second RCP pulse will be associated to a second ECM beam
arising from the northern auroral ring (RCP2).  

The two ECM beams are not arising from the same points of the auroral circle. 
For clarity, in Fig.~\ref{ecm_b} we also included the auroral circles showing the magnetic longitudes $\zeta$
($\zeta=0^{\circ}$ on the the plane identified by the magnetic and rotation axis) 
of the magnetic field lines responsible for the detectable auroral radio emission. 
Such magnetic field lines intercept the northern and the southern auroral rings in the two sectors 
characterized by ECM beam pattern crossing the observer's line of sight.
%
%
In particular, for the specific case analysed here, the magnetic longitudes of the field lines that have the most favourable orientation for
the auroral radio emission to be detected are respectively:
$\zeta \approx 135^{\circ}$ and $\zeta \approx 315^{\circ}$ (peak RCP1); $\zeta \approx 45^{\circ}$ and $\zeta \approx 225^{\circ}$ (peak RCP2);  
$\zeta \approx 125^{\circ}$ and $\zeta \approx 305^{\circ}$ (peak LCP1); $\zeta \approx 55^{\circ}$ and $\zeta \approx 235^{\circ}$ (peak LCP2).


\begin{figure*}
\resizebox{\hsize}{!}{\includegraphics{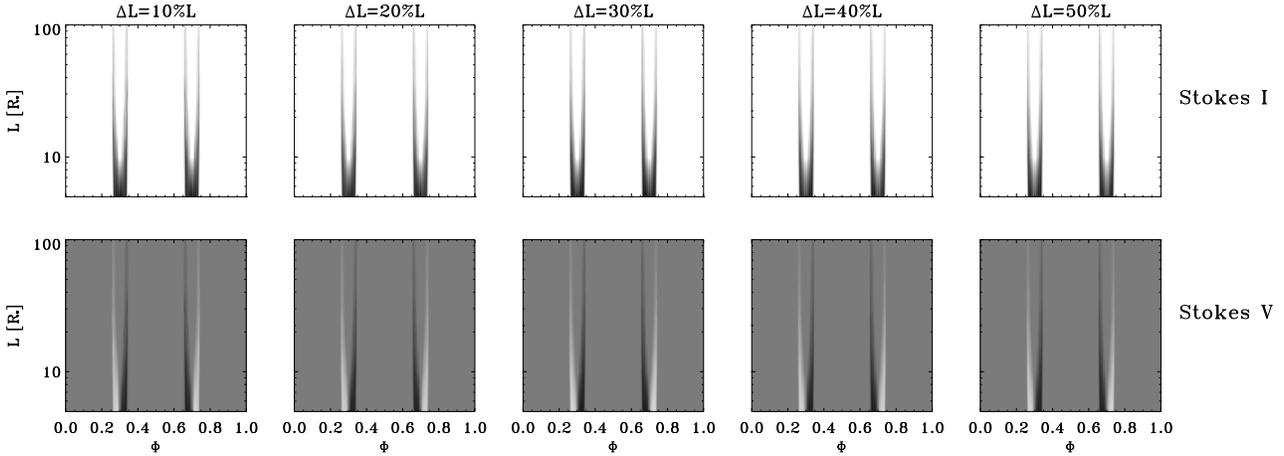}}
\caption{Dynamic auroral light curves as a function of the magnetic shell size.
Top panels: Stokes I; middle panel: Stokes V.
The simulations have been performed varying the equatorial extension
of the magnetospheric cavity where the auroral radio emission takes place (defined by the $L$-shell parameter),
and the corresponding shell thickness.
}
\label{size_thick}
\end{figure*}

\subsection{Dependence on the magnetic shell size}

We performed model simulations of the auroral radio emission arising from magnetic shells with sizes different from the 
one previously analysed. The simulated ECM frequency and the radiation beam pattern were left unchanged
($\nu=1.4$ GHz, $\delta=20^{\circ}$ and $\theta=10^{\circ}$).
We analyse the cases of auroral cavities delimited by the inner boundary magnetic field lines with the $L$-shell parameter
ranging from 5 to 100  R$_{\ast}$. In addition,
the magnetic shell thickness at the equatorial plane has been varied in the range from 10\% to 50\% of the $L$ parameter of the inner 
boundary magnetic field line.

The simulated dynamic ECM light curves (Stokes I and V) as a function of the auroral cavity size have been
displayed in Fig.~\ref{size_thick}. Each column panels of Fig.~\ref{size_thick} show
the simulations performed assuming different thickness of the magnetic cavity. 
Similarly to the previously adopted grey scale for the Stokes V simulations, the white regions
indicate the ECM emission that have a right hand circular polarization sense, whereas the black areas
refer to radiation with left-hand circular polarization.

The result of these simulations highlights
that the size of the auroral cavity significantly affects the pulse profile. 
The simulations of the auroral radio emission for the Stokes I arising from small magnetic shells
are characterized by two large pulses per stellar rotation, see top panels of Fig.~\ref{size_thick}. 
As the size of the magnetic shell increases, 
the two simulated pulses become clearly doubly peaked, each single peak becoming progressively thinner.
Each component is clearly related
to a specific stellar hemisphere, as recognizable from the sign of the simulated light curves for the Stokes V in the middle panels of Fig.~\ref{size_thick}.
The thinning of the pulse with the increasing magnetic shell size is the consequence of the progressive decrease of the auroral ring radius associated to a specific radio frequency
(as extreme case, for a magnetic shell with infinite $L$-shell parameter the auroral region degenerates 
to a point located on the magnetic axis),
and consequently  the number of grid points that are sources of detectable ECM emission
progressively decreases.
This set of simulations highlights that the ECM features  do not depend on the ratio $\Delta L/ L$.



\subsection{Dependence on the stellar geometry}

\begin{figure*}
\resizebox{\hsize}{!}{\includegraphics{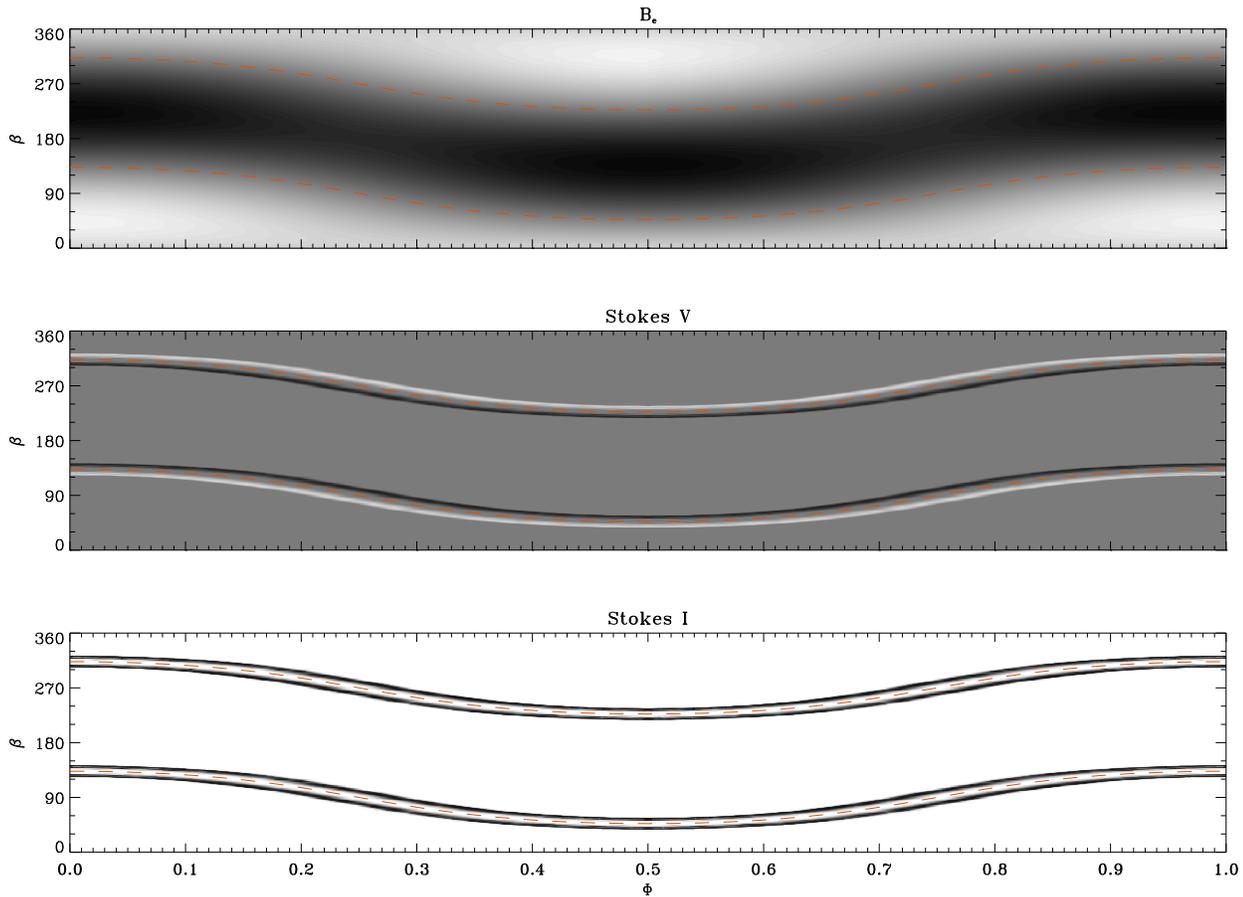}}
\caption{Variations of the effective magnetic field curve and of the auroral light  curve 
as a function of the misalignment between the magnetic and the rotation axis orientations (angle $\beta$).
Top panel: dynamic effective magnetic field curves.
Middle panel: dynamic auroral light curves for the Stokes V.
Bottom panel: dynamic auroral light curves for the Stokes I.
The orange dashed lines indicate the values of  $\beta$ that locate the nulls of the magnetic effective curve 
at fixed values of the stellar rotational phase, obtained using the Eq.~\ref{eq2}.
}
\label{cl_din}
\end{figure*}

To highlight the role of the stellar geometry on the auroral radio emission detectability, 
a set of model simulations have been performed by varying the angle $\beta$.
After setting the auroral shell size ($L=15$ R$_{\ast}$ and $\Delta L= 20\% L$),
we performed simulations at $\nu=1.4$ GHz
(beam pattern defined by $\delta=20^{\circ}$ and $\theta=10^{\circ}$),
varying the magnetic axis obliquity ($\beta$) from $0^{\circ}$ to $360^{\circ}$ (with a simulation step of $1^{\circ}$).
The magnetic polar strength and the rotation axis inclination 
were left unchanged ($B_{\mathrm p}=3000$ Gauss, $i=43^{\circ}$).

The variation of the  $\beta$ parameter significantly affects
the features of the ECM light curves and  the stellar magnetic field curve.
We then compared the simulated ECM light curves with the corresponding effective magnetic field curve.
The dynamic light curves and the magnetic curves are displayed in Fig.~\ref{cl_din}   as a function of $\beta$. 
The effective magnetic field is displayed in the top panel of the figure and the other two panels show the Stokes I and V 
auroral radio emission (respectively, bottom and middle panels of  Fig.~\ref{cl_din}).

By looking at Fig.~\ref{cl_din}, it is evident that the ECM light curve features depend on the adopted stellar geometry.
In particular, the detectability of the auroral radio emission and the phase separation between the two simulated pulses is a function of $\beta$.
When the magnetic axis is aligned with the rotation axis ($\beta=0^{\circ}$ or $\beta=180^{\circ}$)
the effective magnetic field does not vary. 
The effective magnetic field strength has a positive constant value for $\beta=0^{\circ}$ 
(northern hemisphere prevailing), negative value for $\beta=180^{\circ}$ (southern hemisphere prevailing). 
When the magnetic obliquity increases, the effective magnetic field 
changes as the star rotates.
For $\beta=90^{\circ}-i=47^{\circ}$, the magnetic field curve is characterized by one null at $\Phi=0.5$.
When $\beta$ is close to this value,
the light curve of the auroral radio emission is characterized by only one pulse per stellar rotation,
right-hand circularly polarized, centred at $\Phi=0.5$.
As the magnetic obliquity increases, the effective magnetic field inverts its sign twice at every stellar rotation. 
Therefore, we simulated two doubly peaked ECM pulses.

At  $\beta=90^{\circ}$, the pulse separation increases up to $\Delta \Phi=0.5$, where it starts to decrease.
For $\beta=180^{\circ}-i=133^{\circ}$, the magnetic field curve has again a single null, but at $\Phi=0$.
When the misalignment between magnetic and rotation axis is close to this value,
the light curve of the auroral radio emission shows a single large pulse, left-hand circularly polarized and centred at $\Phi=0$.
Beyond this limit of $\beta$, the auroral radio emission vanishes.
The exact limits of the range of $\beta$ able to give detectable auroral radio emission 
also depend on  the ray path deflection $\theta$.
The simulations performed when $\beta$ lies in the range $180^{\circ}$--$360^{\circ}$
closely resemble the behaviour simulated in the range $0^{\circ}$--$180^{\circ}$, 
but with the opposite sign of $B_{\mathrm e}$. Consequently, a swap of circular polarization occurs
(middle and top panel of Fig.~\ref{cl_din}). 
As a result of the present analysis, it can be noticed that the capability to detect the ECM radiation is marginal
for those stars that do not change their net polarity of the effective magnetic field on the observer's line of sight.

The  beam pattern chosen to perform the present analysis 
is up-ward deflected. The simulated ECM radio light curves are then
characterized by contributions from the two opposite hemispheres that are clearly distinguishable 
by the sign of the simulated Stokes V parameter. 
If the ray path is not deflected ($\theta=0$),
the simulated auroral radio emission is unpolarized (Stokes V$=0$).
The Stokes I light curve is instead characterized by two single peaks
occurring at phases ($\Phi _{\mathrm {0}}$ and $1-\Phi _{\mathrm {0}}$),
which represent the stellar orientations characterized by the magnetic axis being perpendicular
to the line of sight ($B_{\mathrm e}=0$), according to the analysis performed in Sec.~\ref{sec3_1}.
In this case, the detection of the auroral radio emission from a star
can be used to  fix the oblique rotator geometry  ($\beta$), the rotation axis inclination $i$ being known. 
In fact, on the basis of the simple dipolar geometry,
from the Eq.~\ref{eq1} the following relation can be derived:

\begin{equation}
\beta = \arctan  \left( - \frac{1} { \tan i \cos \Phi _{\mathrm {0}}}  \right)
\label{eq2}
\end{equation}

Given the rotation axis inclination, the equation above returns those value of $\beta$ for which the effective magnetic field curve
is zero at $\Phi _{\mathrm {0}}$.
The results of the Eq.~\ref{eq2}, obtained assuming $i=43^{\circ}$,
have been superimposed on the model simulation  in Fig.~\ref{cl_din}.
It can be noticed that the curves $\beta$ versus $\Phi_0$ are intermediate
between the tracks left by the characteristic double peaks of opposite polarity, which are a signature of deflected ECM emission
(middle and bottom panels of Fig.~\ref{cl_din}).
Therefore, in the case of auroral radio emission characterized by ray path deflection ($\theta> 0$), it is also possible to obtain the ORM geometry.
This is because if the ECM contributions arising from the two opposite stellar hemispheres are clearly detected, then
the stellar rotational phase related to a null of the effective magnetic field 
can be located in the middle between the phases of occurrence of the coherent peaks with opposite Stokes V sign.

\section{Discussion and Conclusions}


The purpose of the model described in this paper is the simulation of the auroral radio emission from stars
characterized by a dipole-like magnetic field.
The auroral radio emission is a well-known phenomenon common to the magnetized planets
of the solar system \citep{zarka98}, and also present in some kinds of magnetic stars \citep{trigilio_etal11,nichols_etal12}.
The mechanism responsible for this kind of radio emission is the coherent amplification mechanism known as ECM.
The modelling of the ECM emission has been developed          
in accordance
with the laminar source model \citep{louarn_lequeau96a,louarn_lequeau96b}.
In this case, the auroral radio emission is constrained within a narrow beam tangentially directed along the cavity boundary.
This kind of anisotropic beaming
is able to successfully reproduce the timing and the pulse width of the auroral radio emission observed
from the early type star CU\,Vir (SP A0V) \citep{trigilio_etal11,lo_etal12}. Moreover, at the bottom of the main sequence,
the presence of ECM emission characterized by strongly anisotropic beaming has been confirmed 
in the case of the ultra cool dwarfs (SP $>$ M7) \citep*{lynch_etal15}.

The dependence of the auroral radio emission features 
on the model parameters has been extensively analysed.
Despite the simplified assumption of a pure dipole characterizing the stellar magnetic field, 
this model is a powerful tool to study how the timing and the pulse profile 
depend on the parameters that define the source geometry and on
the parameters that control the beam pattern.
The analysis in this paper allows us
to draw some general conclusions that help us to interpret the
features observed in the auroral radio emission from individual stars.

First of all, we point out
how the recurrence phases of the ECM pulses are related to the features of the magnetic curve. 
In the cases of the auroral radio emission propagating perpendicularly to the magnetic
field vector, we obtain unpolarized coherent pulses 
coinciding exactly with the nulls of the effective magnetic field curve,
since RCP and LCP rays have the same direction.
This coincidence ceases as the ECM beam pattern deflects.
In this case, each single auroral radio pulse becomes doubly peaked, because  the two polarizations are deflected differently. As a consequence of the magnetic polarity of
the stellar hemisphere where the ECM emission arises, the peak components are circularly polarized with opposite polarization senses.

In particular, the two components of opposite polarization sense occur
at phases slightly preceding or following the effective magnetic field null.
The phase separation between the two polarized components
can be a function of such model parameters as the ray path deflection,
the beam opening angle, or the magnetic shell size.
In all cases, the two components are symmetric with respect
to the magnetic field null phase.


As a further result of our simulation, we found that once we set
the size of the magnetic shell where the auroral radio emission takes place, 
the distance from the surface, hence the size of the auroral ring,
does not affect the simulated pulse profile and their localization in phase.
It has also been established that the magnetic shell thickness has negligible effects
on the light-curve features. 
Moreover, the size of the magnetic shell does not affect the timing of the auroral pulses,
while it does have an influence on the pulse profile.
In particular, if the auroral radio emission is generated very close to the magnetic dipole axis, as in the case of the
very small auroral rings related to large magnetospheric shells, 
the ECM arising from each stellar hemisphere has a very narrow pulse width.
We can conclude that, once defined the magnetic shell and fixed the beaming of the ECM emission, 
all the auroral radio frequencies that originate from regions located at different heights above the stellar surface
are characterized by similar light curves. 
{About the model parameters analyzed above,
we can conclude that the auroral radio emission features are frequency independent.
In the sense that, 
once fixed the parameters that control the beam pattern,
the location above the poles of the auroral
rings, that are frequency dependent, have no effect on the modeled auroral radio emission.
However, as discussed below, other frequency dependent effects can take place on the ECM stellar radio emission.
Moreover, the ECM beam pattern is reasonably related to the auroral ring geometry,
roughly speaking auroral radio emission detectable in direction strongly misaligned from the plane tangent to
the cavity wall should be generated in large thickness auroral ring,
conversely strongly beamed ECM emission is related to very thin auroral cavity.}

The discovery that
the light curve of the auroral radio emission does not depend on the height 
above the surface of the auroral ring where it originates has a direct implication on
the parameters that are able to locate it.
These parameters 
are the polar magnetic field strength
and the harmonic of the gyrofrequency amplified by the ECM process.
From the analysis performed in this paper, it follows that the above  
parameters cannot be directly deduced only from the detection of the auroral radio emission from a given star.



The measurement of possible frequency-dependent effects of the pulsed auroral radio emission, such as the frequency drift of the pulses' arrival time,
has to be related to a
possible propagation effect suffered by the auroral radio emission generated at different frequencies 
(for example, the ambient thermal medium 
refracts differently the ECM radiation of different frequencies, as proposed by \citealp{trigilio_etal11}),
or to the intrinsic nature of the basic process that generates the ECM amplified radiation 
(for example, 
{in case of the loss-cone driven ECM emission the hollow cone opening angle is a function of the frequency).
In detail, the half-aperture of the hollow cone is defined as follow:
$\cos \theta_{\mathrm B}  = v/c \sqrt{1-\nu_{\mathrm B}/\nu_{\mathrm {B_{max}}}} $ \citep{hess_etal08}, 
where, 
following the resonance condition for the gyromagnetic emission at the $s$-th harmonic, 
$\nu_{\mathrm B}=\nu /s$  \citep{melrose_dulk82},
with 
$\nu_{\mathrm {B_{max}}}$ the gyro-frequency at the stellar surface.
%
%
We note how the elementary process which gives rise to the amplified radiation could affect the propagation direction 
of the resulting auroral radio emission detected at different frequencies.
On the other hand, the ECM radiation, originated in a density depleted region
close to the stellar surface, along its path passes through refracting cold thermal regions 
that deflect the propagation direction. The ray path refraction
is described by the law of Snell: $\sin \alpha_{\mathrm r} =\sin \alpha_{\mathrm i} / n_{\mathrm{refr}}$, 
where $n_{\mathrm{refr}} = \sqrt{1-\nu^2_{\mathrm{p}} / (\nu(\nu - \nu_{\mathrm B}))}$ 
is the refraction index of the cold thermal medium traveled by the amplified rays,
and $\alpha_{\mathrm i}$ and $\alpha_{\mathrm r}$ are the angles that the incident and the refracted 
rays form with the line perpendicular to the refractive layer.
The refraction plane orientation depends on the shape of such refractive layers,
this can introduce a possible further longitudinal deflection to the ray path
giving rise to a possible phase shift of the coherent pulses.
Moreover, the plasma frequency depends from the density of the thermal plasma ($N_{\mathrm e}$) 
as follow: $\nu_{\mathrm p}=9\times 10^{-6} \sqrt{N_{\mathrm e}}$ GHz,}
a possible longitudinal anisotropy of the ambient thermal medium,
trapped within the stellar magnetosphere,
could affect the propagation of the two ECM pulses differently.
We highlight that the comparison among the magnetic curve and the Stokes I and V light curves of the auroral radio emission
can be used as a diagnostic tool of the environment where the ECM is pumped.

Finally, the simulations performed here allow us to highlight that
the main parameter that localizes the auroral radio pulses in the phase-folded light curve is the obliquity of the stellar magnetosphere
with respect to the rotation axis.
As expected, the timing of the auroral radio pulses
is strictly related to the geometry of the star.
Moreover, the magnetic axis orientation is strictly related to the possibility to detect such a phenomenon.
We can conclude that the study of the timing of the ECM pulses from stars that show such
an elusive phenomenon could be a very useful tool
to obtain hints about the geometry of the magnetosphere where it takes place,
besides its extraordinary importance in the study of the physical
conditions able to generate the ECM radiation itself.


\section*{Acknowledgments}
We thank the referee for his/her constructive criticism which enabled us to improve this paper.


\end{document}